\documentclass[prd,reprint,nofootinbib,showpacs,superscriptaddress]{revtex4-1}
\usepackage{graphicx} 
\usepackage{hyperref}
\usepackage{amsfonts}
\usepackage{amsmath,amssymb}
\usepackage{bm} 
\usepackage{color}
\usepackage{epstopdf} 
\usepackage{epsfig}
\usepackage{subfig} 
\usepackage{float}
{\rm }

\def\be{\begin{equation}}
 \def\ee{\end{equation}}
 \def\bea{\begin{eqnarray}}
 \def\eea{\end{eqnarray}}
 \def\bes{\begin{eqnarray}}
 \def\ees{\end{eqnarray}}
 \def\bi{\begin{itemize}}
 \def\ei{\end{itemize}} % ------- Define Greek Lowercase --------

\def\2{\frac{1}{2}}
\def\4{\frac{1}{4}}

\begin{document}

\title{Passive state preparation in the Gaussian-modulated \\
coherent states quantum key distribution}

\author{Bing Qi}
\email{qib1@ornl.gov}
\affiliation{Quantum Information Science Group, Computational Sciences and Engineering Division, Oak Ridge National Laboratory, Oak Ridge, TN 37831, USA}
\affiliation{Department of Physics and Astronomy, The University of Tennessee, Knoxville, TN 37996, USA}

\author{Philip G. Evans}
\affiliation{Quantum Information Science Group, Computational Sciences and Engineering Division, Oak Ridge National Laboratory, Oak Ridge, TN 37831, USA}

\author{Warren Grice}
\affiliation{Quantum Information Science Group, Computational Sciences and Engineering Division, Oak Ridge National Laboratory, Oak Ridge, TN 37831, USA}

\date{\today}
\pacs{03.67.Dd}

\begin{abstract}

In the Gaussian-modulated coherent states (GMCS) quantum key distribution (QKD) protocol, Alice prepares quantum states \emph{actively}: for each transmission, Alice generates a pair of Gaussian-distributed random numbers, encodes them on a weak coherent pulse using optical amplitude and phase modulators, and then transmits the Gaussian-modulated weak coherent pulse to Bob. Here we propose a \emph{passive} state preparation scheme using a thermal source. In our scheme, Alice splits the output of a thermal source into two spatial modes using a beam splitter. She measures one mode locally using conjugate optical homodyne detectors, and transmits the other mode to Bob after applying appropriate optical attenuation. Under normal conditions, Alice's measurement results are correlated to Bob's, and they can work out a secure key, as in the active state preparation scheme. Given the initial thermal state generated by the source is strong enough, this scheme can tolerate high detector noise at Alice's side. Furthermore, the output of the source does not need to be single mode, since an optical homodyne detector can selectively measure a single mode determined by the local oscillator. Preliminary experimental results suggest that the proposed scheme could be implemented using an off-the-shelf amplified spontaneous emission source. \footnote{This manuscript has been authored by UT-Battelle, LLC under Contract No. DE-AC05-00OR22725 with the U.S. Department of Energy. The United States Government retains and the publisher, by accepting the article for publication, acknowledges that the United States Government retains a non-exclusive, paid-up, irrevocable, world-wide license to publish or reproduce the published form of this manuscript, or allow others to do so, for United States Government purposes. The Department of Energy will provide public access to these results of federally sponsored research in accordance with the DOE Public Access Plan (http://energy.gov/downloads/doe-public-access-plan).
}

\end{abstract}

\maketitle

\section{Introduction}
\label{sec:1}

Quantum key distribution (QKD) has drawn a lot of attention for its proven security against adversaries with unlimited computing power \cite{BB84,E91,Gisin02,Scarani09,Lo14,Diamanti16}. In QKD, two remote legitimate users (Alice and Bob) can establish a secure key by transmitting quantum states through an insecure channel controlled by the adversary (Eve). Any attacks by Eve will, with a high probability, disturb the transmitted quantum state, and thus can be detected.

Many practical QKD systems are based on the so-called prepare-and-measure scheme, where Alice prepares quantum states and transmits them to Bob, who in turn performs measurements. The quantum state preparation step is conventionally implemented in an \emph{active} manner: Alice first generates truly random numbers using a quantum random number generator \cite{MY16,HG17}, which she uses to prepare a corresponding quantum state by performing modulations on the output of a single source, or switching among multiple sources. One well-known example is the decoy state BB84 QKD using phase randomized weak coherent sources \cite{Decoy1, Decoy2, Decoy3}, where for each transmission, Alice needs to randomly prepare one of the four BB84 states \cite{BB84}, randomly change the average photon number (to generate either the signal state or one of the decoy states) and (in certain implementations) the global phase of the weak coherent state \cite{ZQ07}. As the transmission rate in QKD has been growing dramatically over the years, it is becoming more and more challenging to prepare quantum state precisely at the corresponding speed.

More recently, \emph{passive} state preparation schemes have been proposed in QKD as an alternative approach \cite{Mauerer07,Adachi07,Ma08,Curty09,Curty10,Curty102,Zhang10,Zhang12,Sun14,Li14,Zhou14,Curty15,Sun16}. In this scheme, Alice explores intrinsic fluctuations of the source, or intentionally designs the source in a way such that certain parameters (for example, intensity) will present unpredictable fluctuations. Typically, two optical modes with correlated fluctuations are output from the source. By measuring one mode locally, Alice can determine the random noise carried by the other mode, which will be transmitted to Bob. This idea was initially proposed as a simple way to generate random intensity fluctuations in the decoy-state QKD protocols \cite{Mauerer07,Adachi07,Ma08}. Later on, it was also applied in preparing the four BB84 states approximately \cite{Curty15}.

So far the passive state preparation scheme has only been studied in discrete-variable (DV) QKD based on single photon detection. Here we propose a passive state preparation scheme in continuous-variable (CV) QKD based on coherent detection \cite{Ralph99, Hillery00, GMCS}.

CV-QKD based on coherent detection could potentially be a cost-effective solution in practice, especially over a relatively short distance \cite{Diamanti15}. One of the most promising CV-QKD protocols is the Gaussian-modulated coherent states (GMCS) QKD \cite{GMCS}, which has been demonstrated over practical distances \cite{GMCS, Lodewyck07, Qi07, Jouguet13, Kumar15, Huang16, ZLC17}. Similar to the case of DV QKD, in the GMCS QKD, quantum states are prepared actively: for each transmission, Alice first generates a pair of Gaussian-distributed random numbers, then encodes them on a weak coherent state using optical amplitude and phase modulators. Since the modulation format is relatively complicated and the tolerable modulation error is small, high extinction ratio modulators with good stability are required in the GMCS QKD \cite{Jouguet12}.

In this paper, we propose a passive state preparation scheme using a thermal source. One observation is that in the GMCS QKD, from Eve and Bob's point of views, the quantum states sent by Alice are thermal. So, instead of preparing a thermal state from a coherent state by preforming Gaussian modulations, Alice can simply use a thermal source: Alice splits the output of the thermal source into two spatial modes using a beam splitter. She measures both the X and P quadratures of one mode using conjugate optical homodyne detectors, and transmits the other mode to Bob after applying appropriate optical attenuation. To estimate the quadrature values of the outgoing mode, Alice can scale down her measurement results by the attenuation applied on the outgoing beam. At Bob's end, he performs similar measurements to determine the quadrature values of the received state. Under normal conditions, Alice's measurement results will be correlated to Bob's, and they can further work out a secure key. Note in this scheme, the shared randomness originates from the intrinsic quadrature fluctuations of a thermal state. We remark that our propsoal is differet from previous studies on CV-QKD using noisy coherent state or thermal state \cite{Filip08, Usenko10, Weedbrook10}, where the super-Poissonian photon statistics of the source is regarded as excess noise in CV-QKD based on active state preparation. It was first proposed in  \cite{Filip08} that the above preparation noise could be suppressed by increasing the modulation variance and then applying strong attenuation.  

One may wonder whether the vacuum noise introduced by Alice's conjugate optical homodyne detection will ultimately prevent her from acquiring a precise estimation of the quadrature values of the outgoing mode. As we will show in this paper, given the initial thermal state generated by the source is strong enough, the contribution of Alice's detector noise on the estimation error of the outgoing state can be reduced effectively by introducing high attenuation on the outgoing mode.

In practice, it may be difficult to prepare a \emph{single mode} thermal state and match its spectral-temporal mode with that of the local oscillator (LO) used in homodyne detection. Fortunately, to implement our protocol, Alice's thermal source does not need to be single mode. The LO in homodyne detection acts as a mode ``filter'' and can selectively measure only one mode emitted by the source. By using a multimode (broadband) source, it is also easy to align the central wavelength of the LO within the spectrum of the thermal source. 

This paper is organized as follows: In Section \ref{sec:2}, we present details of the GMCS QKD based on a passive state preparation scheme. In Section \ref{sec:3}, we conduct numerical simulations to estimate the potential secure key rates of the proposed scheme. In Section \ref{sec:4}, we characterize the output of a practical amplified spontaneous emission (ASE) source. Preliminary results suggest that such a source could be employed to implement the passive state preparation scheme. Finally, we conclude this paper with a discussion in Section \ref{sec:5}.

\section{Passive state preparation scheme}
\label{sec:2}

Presently, the GMCS QKD protocol is implemented based on the prepare-and-measure scheme: for each transmission, Alice draws two random numbers $x_A$ and $p_A$, prepares a coherent state $|x_A+ip_A\rangle$ accordingly, and sends the prepared state to Bob through an insecure quantum channel. Here $x_A$ and $p_A$ are Gaussian random numbers with zero mean and a variance of $V_A N_0$, where $V_A$ is the modulation variance chosen by Alice, and $N_0$ = 1/4 denotes the shot-noise variance. At Bob's end, he can either implement the homodyne protocol \cite{GMCS} by randomly measuring X or P quadrature for each incoming pulse, or he can implement the heterodyne protocol \cite{Weedbrook04} by first splitting the incoming pulse into two, and then measuring one in X basis and the other in P basis. After the above quantum transmission stage, Alice and Bob estimate the transmission efficiency and added noise of the channel. If the observed noise is below a certain threshold, they can work out a secure key by performing reconciliation and privacy amplification.

Note, from Eve and Bob's point of views, the state from Alice is a single mode thermal state with an average photon number of $V_A/2$. In fact, the security of the GMCS QKD is commonly proved based on an equivalent entanglement-based protocol \cite{Grosshans03}, where Alice performs conjugate homodyne detection on one mode of a two-mode squeezed vacuum state and sends the other mode to Bob. In this picture, the state from Alice is indeed thermal.

Here, we propose a passive state preparation scheme using a thermal source. While this protocol can be conveniently implemented with a multi-mode thermal source, for simplicity, in this section we assume the thermal source is single mode. The protocol is summarized as follows (see Fig.1).

\begin{enumerate}
\item Alice splits the output of a thermal source into two spatial modes ($mod_1$ and $mod_2$ in Fig.1) using a 50:50 beam splitter. We assume the average output photon number of the source is $n_0$.
\item Alice attenuates the average photon number of $mod_1$ down to $V_A/2$ by using an optical attenuator and transmits it to Bob. Here $V_A<n_0$ is the desired modulation variance. 
\item Alice measures both the X and P quadratures of $mod_2$ by performing conjugate homodyne detection. From her measurement results of $\lbrace x_2,p_2\rbrace$, Alice estimates the quadrature values of the outgoing mode as $x_A=\sqrt{\frac{2\eta_A}{\eta_D}}x_2$ and $p_A=\sqrt{\frac{2\eta_A}{\eta_D}}p_2$, where $\eta_A$ is the transmittance of the optical attenuator and $\eta_D$ is the efficiency of Alice's detector.
\item Bob measures both the X and P quadratures of the received quantum state by performing conjugate homodyne detection. His measurement results are $\lbrace x_B,p_B\rbrace$.
\item Alice and Bob repeat the above process many times.
\item Alice and Bob perform reconciliation and privacy amplification on the raw data $\lbrace x_A,p_A\rbrace$ and $\lbrace x_B,p_B\rbrace$. Given the observed noise is below a certain threshold, they can work out a secure key. This step is the same as in the active state preparation scheme.
\end{enumerate}

\begin{figure}[t]
	\includegraphics[width=.45\textwidth]{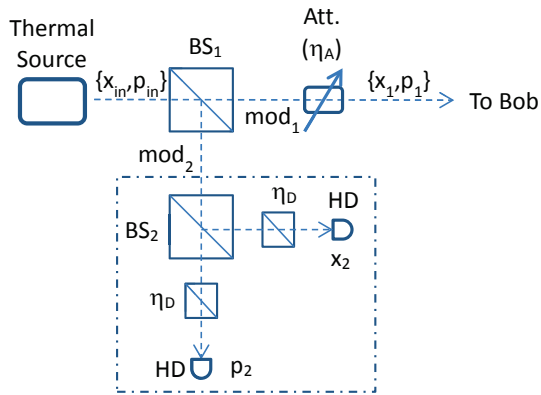}
	\captionsetup{justification=raggedright,
					singlelinecheck=false }
	\caption{The proposed passive state preparation scheme in the GMCS QKD. BS$_1$/BS$_2$-50:50 beam splitter; Att.-optical attenuator; HD-homodyne detector. The efficiency of homodyne detector is modeled by a beam splitter with a transmittance of $\eta_D$. Note the combination of BS$_1$ and Att. could be replaced by an asymmetric beam splitter.} 
	\label{fig:1}
\end{figure}

Note from Eve's point of view, the quantum state sent by Alice in this passive state preparation scheme is the same as the one in the conventional active state preparation scheme. So the well-established security proofs of the GMCS QKD can be applied directly in our scheme. To evaluate the secure key rate, we need to determine how much additional noise will be introduced by this passive state preparation scheme. As we will show below, given the thermal state generated by the source is bright enough, our scheme can tolerate high noise and low efficiency of Alice's detector. We remark that the combination of BS$_1$ and Att. in Fig.1 could be replaced by an asymmetric beam splitter.

For simplicity, we only consider the X-quadrature below. The P-quadrature can be studied in a similar way. The X-quadrature of the outgoing state is given by
\bes\label{eq1} x_1=\sqrt{\frac{\eta_A}{2}}x_{in}+\sqrt{1-\frac{\eta_A}{2}}x_{v1}, \ees 
where $x_{in}$ stands for the X-quadrature of the output of the source, $\eta_A$ is the transmittance of the optical attenuator, and $x_{v1}$ represents vacuum noise introduced by the beam splitter and the attenuator.

Similarly, Alice's measurement result of X-quadrature is given by
\bes\label{eq2} x_2=\sqrt{\frac{\eta_D}{4}}x_{in}+\sqrt{1-\frac{\eta_D}{4}}x_{v2}+N_{el}, \ees 
where $\eta_D$ and $N_{el}$ are the efficiency and noise of Alice's detector; $x_{v2}$ represents vacuum noise due to the two 50:50 beam splitters and the loss of detector. We assume $N_{el}$ is Gaussian noise with zero mean and a variance of $\upsilon_{el}$. In this paper, all of the noise variances are defined in the shot-noise unit.

Alice can estimate $x_1$ from her measurement result $x_2$ using
\bes\label{eq3} x_A=\sqrt{\frac{2\eta_A}{\eta_D}}x_2. \ees 

Using (1) to (3), Alice's uncertainty on $x_1$ is given by
\bes\label{eq4} \Delta=\langle (x_A-x_1)^2 \rangle=\frac{2\eta_A}{\eta_D}(1+\upsilon_{el}-\dfrac{\eta_D}{2})+1. \ees

From (4), the excess noise (the noise above vacuum noise) due to the passive state preparation scheme is given by
\bes\label{eq5} \varepsilon_A=\Delta-1=\frac{2\eta_A}{\eta_D}(1+\upsilon_{el}-\dfrac{\eta_D}{2}). \ees

From (5), by increasing the attenuation on the outgoing mode (decreasing $\eta_A$), the excess noise $\varepsilon_A$ can be effectively reduced. The maximum attenuation Alice can apply is constrained by the average photon number $n_0$ of the thermal state produced by the source and the desired modulation variance $V_A$. Using the relation $V_A=\eta_A n_0$, we can revise (5) as
\bes\label{eq6} \varepsilon_A=\frac{2V_A}{n_0\eta_D}(1+\upsilon_{el}-\dfrac{\eta_D}{2}). \ees

From (6), given a desired modulation variance $V_A$, the brighter the source, the smaller the excess noise introduced by Alice. A typical homodyne detector in the GMCS QKD can achieve $\eta_D=0.5$ and $\upsilon_{el}=0.1$. For a typical value of $V_A=1$, to reduce the excess noise $\varepsilon_A$ below 0.01, the required average photon number of the source is about 340 (per spatial-temporal mode), which can be satisfied by a practical ASE source, as shown in Section IV.

\section{Simulation results}
\label{sec:3}

We conduct numerical simulations of the secure key rates of the passive state preparation scheme. The asymptotic secure key rate of the GMCS QKD, in the case of reverse reconciliation, is given by Refs.~\cite{Lodewyck07, Fossier09} 
\bes\label{eq7} R=fI_{AB}-\chi_{BE}, \ees
where $I_{AB}$ is the Shannon mutual information between Alice and Bob; $f$ is the efficiency of the reconciliation algorithm; $\chi_{BE}$ is the Holevo bound between Eve and Bob. $I_{AB}$ and $\chi_{BE}$ can be determined from the channel loss, observed noises, and other QKD system parameters.

We assume the quantum channel between Alice and Bob is telecom fiber with an attenuation coefficient of $\gamma$. The channel transmittance is given by
\bes\label{eq8} T=10^{\frac{-\gamma L}{10}},\ees
where $L$ is the fiber length in kilometers.

In the case of conjugate homodyne detection, the noise added by Bob's detector (referred to Bob's input) is given by \cite{Fossier09}
\bes\label{eq9} \chi_{het}=[1+(1-\eta_D)+2\upsilon_{el}]/\eta_D,\ees
where we have assumed that Bob's detector has the same performance as Alice's.

The channel-added noise referred to the channel input is given by
\bes\label{eq10} \chi_{line}=\frac{1}{T}-1+\varepsilon_E,\ees
where $\varepsilon_E$ is the excess noise due to Eve's attack. In practice any \emph{untrusted} noise from the QKD system can be included into $\varepsilon_E$. Here, we separate $\varepsilon_E$ into two terms
\bes\label{eq11} \varepsilon_E=\varepsilon_A+\varepsilon_0,\ees
where $\varepsilon_A$ is the excess noise due to the passive state preparation scheme as given in (6). $\varepsilon_0$ represents other sources of untrusted noise.

The overall noise referred to the channel input is given by
\bes\label{eq12} \chi_{tot}=\chi_{line}+\dfrac{\chi_{het}}{T}. \ees

Since both quadratures can be used to generate secure key, the mutual information between Alice and Bob can be determined by
\bes\label{eq13} I_{AB}=log_2\dfrac{V+\chi_{tot}}{1+\chi_{tot}}, \ees
where $V=V_A+1$.

To estimate $\chi_{BE}$, we adopt the realistic noise model where Eve cannot control the loss inside Bob's system, and the detector noise from Bob is assumed to be trusted \cite{GMCS}. This noise model has been widely used in CV-QKD experiments \cite{GMCS, Jouguet13, Kumar15, Qi07, Lodewyck07}. Under this model, the Holevo bound of the information between Eve and Bob is given by Ref.~\cite{Lodewyck07} 
\bes\label{eq14} \chi_{BE}=\sum_{i=1}^2 G\left( \dfrac{\lambda_i-1}{2} \right)  - \sum_{i=3}^5 G\left( \dfrac{\lambda_i-1}{2}\right),  \ees
where $G(x)=(x+1){\rm{log}}_2(x+1)-x{\rm{log}}_2x$.

\bes\label{eq15} \lambda_{1,2}^2=\frac{1}{2} \left[ A\pm \sqrt{A^2-4B} \right], \ees
where
\bes\label{eq16} A=V^2 (1-2T)+2T+T^2 (V+\chi_{line})^2, \ees
\bes\label{eq17}B=T^2(V\chi_{line}+1)^2. \ees
 
\bes\label{eq18} \lambda_{3,4}^2=\frac{1}{2} \left[ C\pm \sqrt{C^2-4D} \right], \ees
where
\begin{equation}
\begin{split}
C=\dfrac{1}{(T(V+\chi_{tot}))^2} [ A\chi_{het}^2+B+1+2\chi_{het} \\
( V\sqrt{B}+T(V+\chi_{line})) +2T(V^2-1)],
\end{split}
\end{equation}

\bes\label{eq20}D=\left( \dfrac{V+\sqrt{B}\chi_{het}}{T(V+\chi_{tot})} \right) ^2. \ees 
\bes\label{eq21} \lambda_5=1. \ees

Simulation parameters are summarized as follows:
$\gamma=0.2$ dB/km, $\varepsilon_0=0.01$, $\upsilon_{el}=0.1$, $\eta_D=0.5$, and $f=0.95$. The modulation variance $V_A$ is numerically optimized at different fiber lengths. 

\begin{figure}[t]
	\includegraphics[width=.5\textwidth]{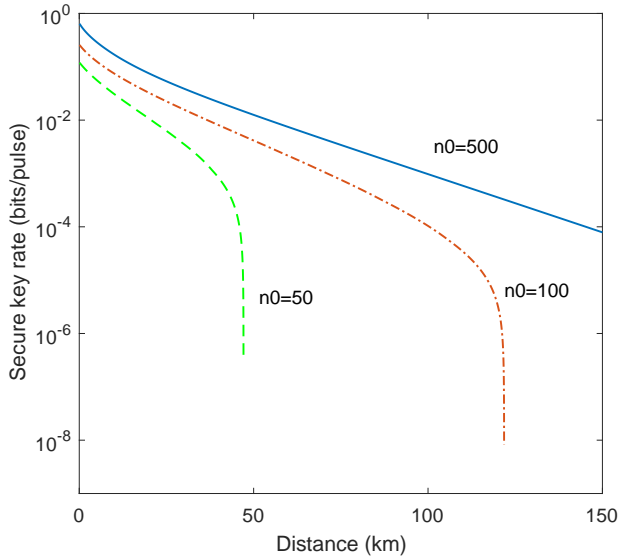}
	\captionsetup{justification=raggedright,
					singlelinecheck=false }
	\caption{Simulation results of the secure key rate for 3 different average photon number of the thermal state: $n_0$=50, 100, 500. Other simulation parameters: $\gamma=0.2$ dB/km; $\varepsilon_0=0.01$; $\upsilon_{el}=0.1$; $\eta_D=0.5$; $f=0.95$}
	\label{fig:2}
\end{figure}

In Fig.~\ref{fig:2} we present the relations of the secure key rate and the fiber length for three different average photon number $n_0$. As shown in Fig.2, a thermal source with an average output photon number above 100 can be employed to implement the passive CV QKD scheme efficiently.  

\section{ASE source characterization}
\label{sec:4}

Previous studies have shown that the ASE noise generated by a fiber amplifier is thermal \cite{WH98, VV00}. In \cite{QL17}, we conducted conjugate homodyne detection and verified the photon statistics of a single mode component (selected by the LO) of an ASE source follows a Bose-Einstein distribution, as expected from a single mode thermal state. Nevertheless, the average photon number of the thermal state in the previous experiment \cite{QL17} is relatively low (about 15). Here, we use a similar setup to characterize the output of a commercial ASE source operated at higher output power.
 
The experimental setup is shown in Fig.3. A fiber amplifier (PriTel, Inc.) with vacuum state input is employed as a broadband thermal source. A 0.8nm optical bandpass filter centered at 1542nm is placed after the ASE source (BP in Fig.3) to reduce the power of unused light. To select out a single polarization mode, a fiber pigtailed polarizer is employed (Pol in Fig.3). A continuous-wave (CW) laser source with a central wavelength of 1542nm (Clarity-NLL-1542-HP from Wavelength Reference) is employed as the LO in coherent detection. Note it is not necessary to stabilize the laser wavelength, which can never drift out the above 0.8nm range under normal operation. A variable optical attenuator (VOA in Fig.3) is used to adjust the LO power, and a fiber polarization controller (PC in Fig.3) is used to match the polarization of the LO with that of the thermal source. To perform conjugate optical homodyne detection, we use a commercial 90$^o$ optical hybrid (Optoplex) and two balanced amplified photodetectors (Thorlabs). The outputs of the two balanced photodetectors are sampled by a real time oscilloscope.

\begin{figure}[t]
	\includegraphics[width=.5\textwidth]{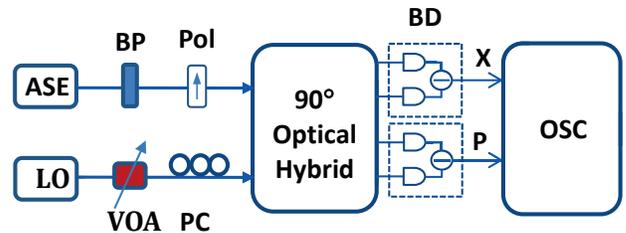}
	\captionsetup{justification=raggedright,
					singlelinecheck=false }
	\caption{Experimental setup. ASE-broadband thermal source; LO-narrow-band laser source; BP-optical band pass filter; Pol-fiber polarizer; VOA-variable optical attenuator; PC-polarization controller; BD-balanced photo-detector; OSC-oscilloscope. } 
	\label{fig:3}
\end{figure}

The noise of the two balanced detectors have been determined to be 0.37 and 0.35 in the shot noise unit. The overall detection efficiency (taking into account the loss of the 90$^o$ optical hybrid and the quantum efficiency of the photo-diode) is about 0.5.
Fig.4 shows the distributions of the measurement results with either a vacuum input or a thermal input. By normalizing the quadrature variances of the thermal state to the vacuum noise, the average photon number (per mode) of the thermal state has been determined to be about 800. As discussed in Section II, such a thermal source is bright enough to implement the passive CV-QKD scheme. Fig.5 shows the 2-dimensional histogram of the measured data when the input is a thermal state. The small deviation from a perfect 2-dimensional Gaussian distribution is most likely due to the non-uniform bin size of the 8-bit analog-to-digital converter of the oscilloscope.

\begin{figure}[t]
	\includegraphics[width=.5\textwidth]{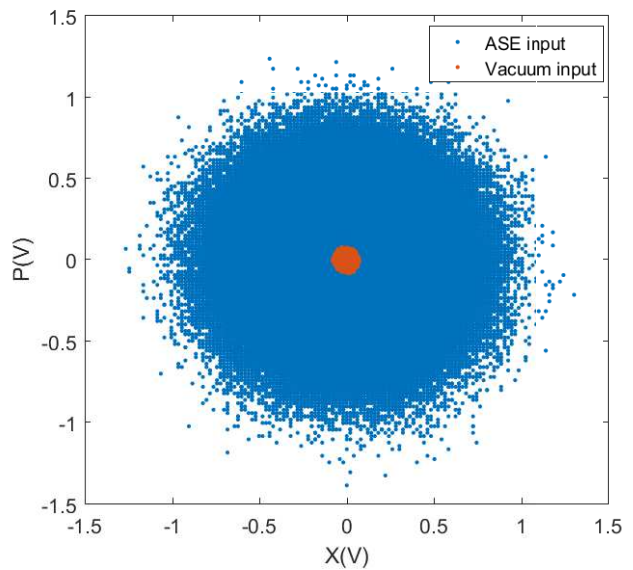}
	\captionsetup{justification=raggedright,
					singlelinecheck=false }
	\caption{Experimental results shown in phase space. The measurement results with both a vacuum input and a thermal state input are presented.} 
	\label{fig:4}
\end{figure}

\begin{figure}[t]
	\includegraphics[width=.5\textwidth]{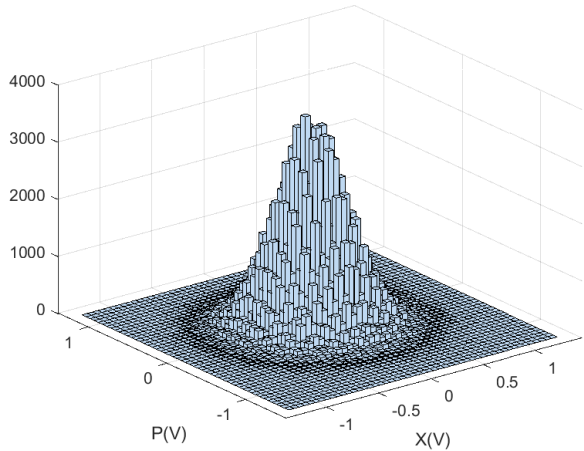}
	\captionsetup{justification=raggedright,
					singlelinecheck=false }
	\caption{The histogram of the experimental results with a thermal state input.} 
	\label{fig:5}
\end{figure}

The single-time second-order correlation function $g^{(2)}(0)$ is an important parameter to characterize a photon source \cite{Glauber63}. In \cite{QL17}, we have shown that $g^{(2)}(0)$ can be conveniently calculated from the statistics of the conjugate homodyne measurement using 
\bes\label{eq21} g^{(2)}(0)=\dfrac{\langle Z^2\rangle-4\langle Z\rangle+2}{(\langle Z\rangle-1)^2},  \ees
where $Z=X^2+P^2$.

From the experimental data, the $g^{(2)}(0)$ of the ASE source has been determined to be 2.012, which is reasonably close to the theoretical value of 2 for a perfect thermal source.

\section{Discussion}
\label{sec:5}

Quantum state preparation is a crucial step in QKD. In the GMCS QKD, this step is implemented using a random number generator, a weak coherent source, and high performance optical modulators. In this paper, we propose a passive state preparation scheme where Alice and Bob generate shared randomness by measuring correlated thermal states split from a common thermal source. This scheme may significantly simplify the implementation of CV-QKD and make it more practical. Note this scheme is different from the entanglement-based QKD where each of Alice and Bob measures a sub-system of an entangled state. In the former, the trustworthiness of the source is required, while in the latter, the entanglement source may be controlled by Eve without compromising the security.

A few more words about imperfections at Alice's side. There are two types of imperfections to be addressed. One is associated with the thermal source itself, and the other is associated with Alice's detector.

To deal with the imperfection of the thermal source itself, precise quantum state tomography can be performed to quantify the deviation of the output of the source from a perfect thermal state. This is especially important when a multi-mode thermal source is employed. We need to make sure there is no correlation between quadrature values of different modes. Otherwise, Eve may gain information without introducing noise by measuring modes not detected by Alice and Bob. Once the imperfection of the source has been quantified, it should be taken into account in the secure proof and key rate calculation. The output of a practical light source may drift with time, which implies that the above state tomography process may need to be repeated time by time. Of course, in the conventional active state preparation scheme, both the laser source and the modulators need to be calibrated over time for similar reasons. It seems reasonable to believe that a thermal source operated in CW mode will show better stability than optical modulators operated at high speed. How to deal with the source imperfections in CV-QKD has drawn much attention \cite{Usenko10, Shen11, Jouguet12, Yang12, Usenko16, Derkach17}. We leave this as a future research topic.

To deal with the imperfections of Alice's detector, in this paper we have made a conservative assumption that Alice's detector noise is untrusted and thus the corresponding excess noise is attributed to Eve's attack. In practice Eve cannot access Alice's system, so one could assume that the above noise is trusted, just like the noise from Bob's detector. Under reverse reconciliation, the trusted noise from Alice's detector will reduce the mutual information $I_{AB}$ but will not change Eve's information $\chi_{BE}$. This trusted noise model can tolerate higher detector noise and work with a lower photon number of the source. We remark that to justify the trusted noise model in practice, specially designed monitoring system may be required to prevent Eve from manipulating the detector performance.    

Note in our scheme, the randomness is generated from a thermal source. Can we trust this randomness? As has been discussed in \cite{Qi17}, while quantum randomness is ultimately connected to quantum superposition states, in the trusted device scenario, the state received by the detector does not need to be a pure state. For example, trusted randomness can be generated by measuring only one photon from an entangled photon pair, given Eve cannot access the other photon. In our scheme, photons from the thermal source are generated through spontaneous emission processes and are entangled with the atoms inside the source. Given the source itself is protected from Eve, true randomness can be generated.

While our theoretical discussions are based on a single mode thermal state, in practice, the proposed scheme can be implemented using a broadband source operated in CW mode (a multi-mode source), thanks to the mode-selective feature of coherent detection. Since the randomness carried by different modes of the source are independent of each other, to generate shared randomness it is crucial that Alice and Bob's detectors measure the same mode of the source. Furthermore, Alice and Bob need a scheme to establish a phase reference between their homodyne detection. Two schemes have been developed in CV QKD based on active state preparation, and both of them can be adopted in the passive state preparation protocol. In this first scheme, Alice generates two strong LOs from the same laser, uses one of them in her local measurement, and sends the second one to Bob to be used as a LO in his measurement. In the second scheme \cite{Qi15,Soh15}, both Alice and Bob generate LOs from their own local lasers. The phase relation between their measurement bases can be determined by sending relatively weak phase reference pulses from Alice to Bob. 

This work was performed at Oak Ridge National Laboratory (ORNL), operated by UT-Battelle for the U.S. Department of Energy (DOE) under Contract No. DE-AC05-00OR22725. The authors acknowledge support from DOE Technology Commercialization Fund and ORNL Technology Transfer and Economic Development (Partnerships) Royalty Funds. 

\textbf{Note added}: During the preparation of this manuscript, we noticed a related work by E. Newton, et al. \cite{Newton17}.

\end{document}